\documentclass[12pt, a4paper]{extarticle}
\usepackage[margin=2cm]{geometry}
\usepackage[affil-it]{authblk}

\usepackage[english]{babel}
\usepackage{graphicx} % Required for inserting images
\usepackage{wrapfig}
\usepackage{amsmath,amsthm,amssymb,scrextend}
\usepackage{booktabs}            % professional-quality tables
\usepackage{multirow}            % tabular cells spanning multiple rows
\usepackage{amsfonts}            % blackboard math symbols
\usepackage{multicol}
\usepackage[shortlabels]{enumitem}
\usepackage{cutwin}
\usepackage{float}
\usepackage{csquotes}
\usepackage{enumitem}
\usepackage{tikz}
\usepackage{hyperref}

\def \bf#1 {\textbf{#1 }}
\def \sumt {\sum\limits}
\providecommand{\keywords}[1]
{
  \small	
  \textbf{\textit{Keywords:}} #1
}

\renewenvironment{proof}{\begin{addmargin}[1em]{0em}\begin{newproof}}{\end{newproof}\end{addmargin}\qed}
\newtheorem{thm}{Theorem}
\newtheorem{lm}{Lemma}
\newtheorem{cor}{Corollary}
\newtheorem*{lm*}{Lemma}

\newtheorem*{defin*}{Definition}

\def \sumt {\sum\limits}

\DeclareMathOperator{\BC}{BC}
\DeclareMathOperator{\Clo}{Clo}

\DeclareMathOperator{\Str}{Str}
\DeclareMathOperator{\cov}{cov}
\DeclareMathOperator{\var}{var}
\DeclareMathOperator{\E}{\mathbb{E}}
\def \cN {\mathcal{N}}

\begin{document}

\title{Theoretical relations between basic centrality measures in real-world networks}
\author{Mikhail Tuzhilin%
  \thanks{Affiliation: HSE university, Moscow State University, Electronic address: \texttt{mikhail.tuzhilin@math.msu.ru};}
}
\date{}
\maketitle

\begin{abstract}
We establish new relationships among several fundamental measures of real-world networks, including the average clustering coefficient, the global clustering coefficient, the average shortest-path length and degree centrality, closeness centrality, betweenness centrality and ksi-centrality. We further develop the theory of $j$-neighborhood centrality, which encompasses degree centrality and ksi-centrality as special cases.
\end{abstract}


%------------------------------------------------------------------

\section{Introduction}

The structural analysis of complex networks is one of the central topics in modern graph theory and network science. Understanding the structure of networks in real-world systems requires the use of quantitative characteristics that capture different aspects of vertex importance, local connectivity, clustering, and global distances.

One of the most widely used structural characteristics are vertex centrality measures. Centrality measures describe the importance of vertices from different perspectives, such as the number of direct connections, the ability to mediate communication between other vertices, or the proximity to the remaining vertices of the network. A wide variety of centrality measures have been proposed in a network science. For example, Shvydun \cite{shvydun} described more than 200 known centrality measures, while his website currently lists over 400. Some of the most widely used centrality measures include degree centrality, local clustering, closeness centrality, and betweenness centrality. The systematic study of these concepts in social network analysis goes back to the classical works of Freeman\cite{free1, free2}.

In real-world networks (constructed from empirical data) the most important characteristics are degree distribution, ksi-distribution, the average clustering coefficient, and the average shortest-path length. The average shortest-path length is one of the key measures used to characterize the small-world properties of a network, whereas the average clustering coefficient reflects the density of connections among neighboring vertices and thus the tendency of the network to form tightly connected communities~\cite{watts}. Another direction concerns the distributions of degree centrality and ksi-centrality. The skewnesses of these distributions are invariants of real networks and distinguish real-world networks from artificially generated ones. In~\cite{tuzjneir}, the author showed that degree centrality and ksi-centrality can be unified within a single sequence of centrality measures called $j$-neighborhood centralities. In this context, Laplacian-based $j$-neighborhood centralities offer an additional approach to describing the local structure of a graph. In particular, the ksi-centrality considered in this work is defined through the Laplacian quadratic form evaluated at the characteristic vector of the open neighborhood of a vertex. 

This article builds on the author's earlier efforts, presented in~\cite{tuzrelL, tuzrelC}, to establish rigorous combinatorial proofs of the relationships between various centrality measures and other graph invariants, and further develops this line of research. In this article we generalized results about the relation between average clustering coefficient and global clustering coefficient, and obtained a new relation between average clustering coefficient and average shortest path length for any graph. In addition, we found many new relations between these values, another centralities and $j$-neighborhood centralities (including degree and ksi-centralities).

\section{$J$-neighborhood centrality and its properties}

Here we propose new relations of $j$-neighborhood centrality from~\cite{tuzjneir}. This centrality is a set of centralities $\xi_i^j$ for a vertex $i$ parametrized by $j$:
$$
    \xi^j_{i} = \frac {\Big(L\chi^j_i, \chi^j_i\Big)} {\big(\chi^j_i, \chi^j_i\big)},
$$
where $\chi^j_i(k)$ is the characteristic function of $\cN^j(i)$, that is
$$
\chi^j_i(k) = \begin{cases} 
1, & k\in \cN^j(i), \\ 
0, & \text{otherwise}.
\end{cases}
$$

For $j = 0$ and $1$ this centrality coincides with degree and ksi-centrality respectively. For larger $j$ the skewness of its distribution is a new invariant of real networks, that differentiates real networks from synthetic ones. Let's use the notations of $j$-neighborhood $\cN^j(i)$ of a vertex $i$ and of the set of edges $E(A,B)$ from the set $A$ to $B$ from~\cite{tuzjneir}.

\begin{lm}\label{lm:jneir}
    Let's fix some vertex $i$. For $j$-neighborhood centrality holds
    \begin{enumerate}
        \item 
        $$
            \xi^j_i = \frac {\{\text{number of shortest paths from $i$ to all vertices of length $j$}\}+\{\text{the same  of length $j+1$}\}} {|\cN^j(i)|},
        $$
        \item
        $$
            1+\frac {|\cN^{j+1}(i)|}{|\cN^{j}(i)|}\leq\xi^j_i\leq |\cN^{j-1}(i)|+|\cN^{j+1}(i)|.
        $$
    \end{enumerate} 
\end{lm}
\begin{proof}
    Let's note, that 
    $$ \xi^j_i\  |\cN^{j}(i)|= E\big(\cN^{j}(i), V\setminus\cN^{j}(i)\big)= E\big(\cN^{j-1}(i),\cN^{j}(i)\big)+E\big(\cN^{j}(i),\cN^{j+1}(i)\big).$$ 
    The first holds the fact, that there is a bijection between the number of shortest paths from $i$ to all vertices of length $j$ and edges of $E\big(\cN^{j-1}(i),\cN^{j}(i)\big)$.
    
    The second holds from two facts: the maximum number of edges $E\big(\cN^{j-1}(i),\cN^{j}(i)\big)$ is $|\cN^{j-1}(i)|\cdot|\cN^{j}(i)|$ and for any vertex of $\cN^{j}(i)$ there exists a least one adjacent edge from $E\big(\cN^{j-1}(i),\cN^{j}(i)\big)$. Therefore, holds
    $$
        \cN^{j}(i)\leq E\big(\cN^{j-1}(i),\cN^{j}(i)\big)\leq |\cN^{j-1}(i)|\cdot|\cN^{j}(i)|.
    $$
    The same for $E\big(\cN^{j}(i),\cN^{j+1}(i)\big)$.
\end{proof}

This lemma allows us to define normalized versions of $j$-neighborhood centralities in two different ways:
\begin{enumerate}
    \item the first way
    $$
        \hat\xi^j_i = \frac {E\big(\cN^{j}(i), V\setminus\cN^{j}(i)\big)} {|\cN^{j}(i)| \big(|\cN^{j-1}(i)|+ |\cN^{j+1}(i)|\big)},
    $$
    \item and the second 
    $$
        \hat\xi^j_i = \frac {E\big(\cN^{j}(i), V\setminus\cN^{j}(i)\big)} {|\cN^{j}(i)| \big(n-|\cN^{j}(i)|\big)},
    $$
    since $|\cN^{j-1}(i)|+ |\cN^{j+1}(i)|\leq n-|\cN^{j}(i)|,$ where $n$ is the number of nodes. Note that in~\cite{tuzcap} the second approach is used.
\end{enumerate}

\begin{lm}\label{lm:dist}
    Let $i$ be a vertex of a simple undirected graph:
    \begin{enumerate}
        \item $
    \sumt_{t\in V(G)}dist(i,t)\geq\sumt_{j=1}^{k}j(\xi^{j}_i-1)(\xi^{j-1}_i-1)...(\xi^1_i-1),\, 
    $ where $k=1,2,\,...\,,e(i)$,
    \item
    $$\hspace{-40pt}
        \sumt_{t\in V(G)}dist(i,t)\leq D(n-1)-\sumt_{j=1}^{k} (D-j) \Big(\xi^{j-1}_i-(\xi^{j-3}_i-1)(\xi^{j-4}_i-1)...(\xi_i^1-1)\xi_i^0\xi_i^{-1}\Big), 
    $$ where $k=1,2,\,...\,,\min\big(e(i),D-1\big),\ \xi_i^{-1} = 1$, and 
    $$
    \xi^{j-1}_i-(\xi^{j-3}_i-1)(\xi^{j-4}_i-1)...(\xi_i^1-1)\xi_i^0\xi_i^{-1} = 
    \begin{cases}
        \xi^0_i, & \text{for } j = 1, \\
        \xi^1_i-\xi^{-1}_i, & \text{for } j = 2, \\
        \xi^2_i-\xi^{0}_i\xi^{-1}_i, & \text{for } j = 3. 
    \end{cases}
    $$
    \end{enumerate}
\end{lm}
\begin{proof}
    Let's note that
    $$
        \sumt_{t\in V(G)} dist(i, t) =\sumt_{j=1}^Dj\,|\cN^j(i)| =\sumt_{j=1}^{D}j\,|\cN^j(i)|+D\big(n-1-\sumt_{j=1}^{D-1}|\cN^j(i)|\big) = D(n-1)-\sumt_{j=1}^{D-1}(D-j)\,|\cN^j(i)|.
    $$
    Therefore,
    $$
        \sumt_{j=1}^lj\,|\cN^j(i)|\leq\sumt_{t\in V(G)} dist(i, t)\leq D(n-1)-\sumt_{j=1}^{k}(D-j)\,|\cN^j(i)|,
    $$ for $l = 1, 2, ...\,e(i),$ and $k=1,2,\,...\,,\min\big(e(i),D-1\big).$
    By the previous lemma we have
    $$
        |\cN^{j}(i)|\leq (\xi^{j-1}_i-1)|\cN^{j-1}(i)|, \text{ and}
    $$
    $$\hspace{-110pt}
        |\cN^{j}(i)|\geq \xi^{j-1}_i-|\cN^{j-2}(i)|\geq\xi^{j-1}_i-(\xi^{j-3}_i-1)|\cN^{j-3}(i)|\geq ....
    $$
    $$\hspace{110pt}
        ... \geq\xi^{j-1}_i-(\xi^{j-3}_i-1)(\xi^{j-4}_i-1)...(\xi^1_i-1)\xi_i^0. 
    $$
\end{proof}

\begin{lm}\label{lm:cycle}
    For any simple graph $G$ and for any $j=1,2,...\,,\min\limits_{i\in V(G)}e(i).$
    $$
        \sumt_{i\in V(G)} \xi^j_i|\cN^j(i)| \geq \sumt_{i\in \cN^j(i)}|\cN^j(i)|d_i-2(2j+1)\,C_{2j+1}(G),
    $$ where $C_{2j+1}(G)$ is the number of $2j+1$-cycles in the graph. 
\end{lm}
\begin{proof}
    Let's note that
    $$
        \xi^j_i|\cN^j(i)| = \sumt_{i\in \cN^j(i)}d_i-2\big|E(\cN^j(i))\big|.
    $$ For a fixed $i$ each edge $E\big(\cN^j(i)\big)$ corresponds to different $2j+1$-cycle in $G$. Therefore, $C_{2j+1}(G)\geq(2j+1)\sumt_{i\in V(G)}E\big(\cN^j(i)\big).$ 
\end{proof}

\begin{lm}
    For any simple graph
    $$
        \sumt_{i\in V(G)} \sumt_{j=0}^{D} \xi^j_i |\cN^j(i)| = 2\sumt_{i\in V(G)}\Str(i), 
    $$where $|\cN^j(i)|=0$, if there is no vertices $k: dist(i,k) = j$.
\end{lm}
\begin{proof}
    By the lemma~\ref{lm:jneir}:
    $$
        \sumt_{j=0}^{D} \xi^j_i |\cN^j(i)| = 2\,\{\text{total number of shortest paths from $i$}\}.
    $$
\end{proof}

\section{Relations for centralities important  for real networks} 
In real networks there four important centralities which related to their structural properties. They are local clustering, radiality, degree and ksi centralities. In this section, we establish new relations between them and other centralities.

Let's propose the requiring centralities and another characteristics for a graph $G$:
\begin{enumerate}
    \item degree $d_i, i\in V(G)$,
    \item let's denote $s_i = d_i(d_i-1), i\in V(G)$,
    \item {local cluster coefficient} $c_i = \frac {\text{number of edges in }\cN^1(i)} {\text{maximum possible number of edges in }\cN^1(i)}= \frac{2 |E(\cN^1(i))|} {d_i(d_i-1)}$,
    \item {average clustering coefficient} $C_{WS} = \frac 1 {n} \sumt_{i\in V(G)} c_i = \frac 1 n \sumt_{i\in V(G)} \frac {\sumt_{j, k\in V(G)} a_{ij} a_{jk} a_{ki}} {d_i (d_i-1)}$,
    \item {global clustering coefficient} $C = \frac {\text{number of closed triplets in $G$}} {\text{number of all triplets in $G$}} = \frac {\sumt_{i, j, k\in V(G)} a_{ij} a_{jk} a_{ki}} {\sumt_{i\in V(G)} d_i (d_i-1)}$,
    \item {vertex eccentricity} $e(i) =\max\limits_{t\in V(G)}d(i,t)$,
    \item {diameter} $D = \max_{s,t\in V(G)} dist(s,t)$,
    \item {average shortest path length} $L = \frac 1 {n(n-1)} \sumt_{s,t\in V(G): s\neq t} dist(s,t)$,
    \item {closeness centrality} $\Clo(i) = \frac {n-1} {\sumt_{t\in V(G)} dist(i, t)}$, 
    \item {stress} $\Str(i) = \sumt_{s,t\in V(G):\, s\neq t\neq i} \sigma_{st}(i)$, where $\sigma_{st}(i)$ is the total number of shortest paths from $s$ to $t$ which contains vertex $i$,
    \item {betweenness} $\BC(i) = \sumt_{s,t\in V(G):\, s\neq t\neq i} \frac {\sigma_{st}(i)} {\sigma_{st}}$, where $\sigma_{st}$ is the total number of shortest paths from $s$ to $t$.
\end{enumerate}

\begin{thm}\label{thm:ccws}
   For average and global clustering in simple undirected graph hold
   $$
    C-C_{WS} = \frac {\cov(s, c)} {\E(s)},
   $$
   where expected value $\E(s) = \frac  1 n\sumt_{i\in V(G)}s_i$ and covariance $\cov(s,c) = \E\Big((s-\E(s))(c-\E(c)\Big)$.
\end{thm}
\begin{proof}
    Note that $\sumt_{j, k\in V(G)} a_{ij} a_{jk} a_{ki} = s_i c_i$, thus $C = {\E(s c)} /{\E(s)}$. Therefore,
    $$
        \frac {\cov(s,c)} {\E(s)} = \frac {\E(sc)} {\E(s)}-\E(c) =  C-C_{WS}.
    $$
\end{proof}

From this theorem we obtain
\begin{cor}
    For any simple undirected graph
    \begin{enumerate}
        \item if $\cov(s,c) > 0$, then $C>C_{WS}$,
        \item if $\cov(s,c) < 0$, then $C<C_{WS}$.
    \end{enumerate}
\end{cor}
This corollary generalizes the result in~\cite{tuzrelL}, since if both $s$ and $c$ are monotonically increasing, covariance $\cov(s,c)>0$ and vice versa: if both $s$ and $c$ are monotonically decreasing, covariance $\cov(s,c)<0$.

\begin{cor}
    For any simple undirected graph
    $$
        \big|C-C_{WS}\big|\leq \frac {\sqrt{\var(s)\var(c)}}{\E(s)}.
    $$
\end{cor}

This corollary shows that the standard deviations of $s$ and local clustering bound the difference between the global clustering and average clustering coefficients in graphs. 

Let's note that in the lemma~\ref{lm:cycle} for the case $j=1$ inequality becomes equality and thus we obtain corollary about a relation between ksi-centrality and the number of triangles in a graph.
\begin{cor}\label{cor:dksi}
    For any simple undirected graph $G$
    $$
        \sumt_{i\in V(G)} \xi_id_i = \sumt_{i\in V(G)}d_i^2-6\,T(G),
    $$
    where $T(G)$ is the number of triangles or closed triplets in $G$.
\end{cor}

This corollary shows that ksi centrality only depends on the number of triangles, if the distribution of degrees is fixed. Thus, the more triangles in a graphs with the same distribution of degrees, the more ksi-centrality. Also, the edge switching operation (see~\cite{eswitch} for edge switching) can be used   for controlling ksi centrality distribution in a graph.

Let's use the lemma~\ref{lm:dist} to find bounds for different centralities in terms of $j$-neighborhood centrality.  

\begin{cor}
    For any simple undirected graph we have a sequence of inequalities
    \begin{enumerate}
        \item 
        $$\hspace{-10pt}
            \Clo(i)\geq \frac {n-1} {D(n-1)-\sumt_{j=1}^{\min\big(D-1,e(i)\big)} (D-j) \Big(\xi^{j-1}_i-(\xi^{j-3}_i-1)(\xi^{j-4}_i-1)...(\xi_i^1-1)\xi_i^0\xi_i^{-1}\Big)}\geq...
        $$
        $$
            ...\geq\frac {n-1} {D(n-1)-\sumt_{j=1}^{k} (D-j) \Big(\xi^{j-1}_i-(\xi^{j-3}_i-1)(\xi^{j-4}_i-1)...(\xi_i^1-1)\xi_i^0\xi_i^{-1}\Big)}\geq ...
        $$
        $$
            ...\geq\frac {n-1}{D(n-1)-(D-1)d_i-(D-2)\xi_i}\geq\frac {n-1}{D(n-1)-(D-1)d_i}\geq \frac 1 D,
        $$
        \item 
        $$
            \Clo(i)\leq \frac{n-1} {\sumt_{j=1}^{e(i)}j(\xi^{j}_i-1)(\xi^{j-1}_i-1)...(\xi_i^1-1)}\leq ...\leq \frac{n-1} {\sumt_{j=1}^{k}j(\xi^{j}_i-1)(\xi^{j-1}_i-1)...(\xi^1_i-1)}\leq ... \leq \frac {n-1} {\xi_i-1}.
        $$
    \end{enumerate}
    
\end{cor}

In particular, we obtain a boundaries for closeness centrality in terms of diameter, ksi and degree for graphs with $D\geq 2$:
$$\frac {n-1} {D(n-1)-(D-1)d_i-(D-2)\xi_i}\leq\Clo(i)\leq \frac {n-1}{\xi-1}.$$

The next two theorems gives new upper and lower bounds on the average shortest path length in terms of averages of $j$-neighborhood centralities and global clustering coefficients.

\begin{thm}
    For any simple undirected graph $G$ with $D\geq 4$ we have a sequence of inequalities
    $$
        \frac{\Xi^1-1} {n-1}\leq L(G)\leq D- \frac {(D-2)(\Xi^1+\overline{d}-1)+(D-3)\Xi^2+(D-4)\Big(\Xi^3-\overline{d^2}+\frac {6T(G)} {n}\Big) }{n-1}\leq
    $$ 
    $$
    \leq  D- \frac {2\overline{d}+(D-2)(\Xi^1-1)+(D-3)\Xi^2} {n-1} \leq D- \frac {(D-1)\overline{d}+(D-2)(\Xi^1-1)} {n-1}\leq D-\frac {(D-1)\overline{d}} {n-1},
    $$
    where $\overline{d}$ is the average degree, $\overline{d^2}$ --- mean square degree, $\Xi^j$ --- average $j$-neighborhood coefficient and $T(G)$ --- the number of triangles in the graph. For smaller diameters $D(G)=1,2,3$ truncated sequences can be used. 
\end{thm}
\begin{proof}
    Let's use the lemma~\ref{lm:dist}:
    \begin{enumerate}
        \item $
        L(G) =  \frac 1 {n(n-1)} \sumt_{s,t\in V(G): s\neq t} dist(s,t) \geq \frac {\sumt_{s\in V(G)}(\xi_s-1)} {n(n-1)} = \frac {\Xi^1-1} {n-1},
        $
        \item $
        L(G)\leq D-\sumt_{j=1}^{1}\frac {D-j}{n(n-1)} \Big(\xi^{j-1}_i-(\xi^{j-3}_i-1)(\xi^{j-4}_i-1)...(\xi_i^1-1)\xi_i^0\xi_i^{-1}\Big) = D-\frac {(D-1)\overline{d}} {n-1},
        $
        \item $
        L(G)\leq D-\sumt_{j=1}^{2}\frac {D-j}{n(n-1)} \Big(\xi^{j-1}_i-(\xi^{j-3}_i-1)(\xi^{j-4}_i-1)...(\xi_i^1-1)\xi_i^0\xi_i^{-1}\Big) = D-\frac {(D-1)\overline{d}+(D-2)(\Xi^1-1)} {n-1},
        $
        \item $
        L(G)\leq D-\sumt_{j=1}^{3}\frac {D-j}{n(n-1)} \Big(\xi^{j-1}_i-(\xi^{j-3}_i-1)(\xi^{j-4}_i-1)...(\xi_i^1-1)\xi_i^0\xi_i^{-1}\Big) = 
        $
        $$
            =D-\frac {(D-1)\overline{d}+(D-2)(\Xi^1-1)+(D-3)(\Xi^2-\overline{d})} {n-1} = D-\frac {2\overline{d}+(D-2)(\Xi^1-1)+(D-3)\Xi^2} {n-1},
        $$
        \item $
        L(G)\leq D-\sumt_{j=1}^{4}\frac {D-j}{n(n-1)} \Big(\xi^{j-1}_i-(\xi^{j-3}_i-1)(\xi^{j-4}_i-1)...(\xi_i^1-1)\xi_i^0\xi_i^{-1}\Big) = 
        $
        $$
         = D-\frac {2\overline{d}+(D-2)(\Xi^1-1)+(D-3)\Xi^2+(D-4)(\Xi^3-\overline{d^2}+\frac {6T(G)} {n}+\overline{d})} {n-1} = 
        $$
        $$
            = D-\frac {(D-2)(\Xi^1+\overline{d}-1)+(D-3)\Xi^2+(D-4)\Big(\Xi^3-\overline{d^2}+\frac {6T(G)} {n}\Big)} {n-1}.
        $$ Here we used the relation from the corollary~\ref{cor:dksi}.
    \end{enumerate}
\end{proof}

By this theorem we obtain bounds for the average shortest path length for the graphs with $D\geq2$ in terms of diameter, average degree and average ksi-coefficient:
$$
    \frac{\Xi-1} {n-1}\leq L(G)\leq D- \frac {(D-1)\overline{d}+(D-2)(\Xi-1)} {n-1}.
$$

\begin{thm}
    For any simple undirected graph 
    $$
        L\geq 1+ \frac {(1-C) \E(s)} {(n-1)d_{max} },
    $$ where $d_{max}$ is the maximum degree.
\end{thm}
\begin{proof}
    Let's use denotations as in~\cite{tuzrelC}. Consider the local betweenness $BC\big(i,\cN^1(i)\big) = BC(i) = \sumt_{s,t\in \cN^1(i):\, s\neq t\neq i} \frac {\sigma_{st}(i)} {\sigma_{st}},$ where shortest paths between vertices $s,t$ are calculated in the whole graph $G$. For average local betweenness holds
    $$
        \sumt_{i\in V(G)} BC\big(i,\cN^1(i)\big) =  \sumt_{i\in V(G)} \sumt_{\substack{s,t\in \cN^1(i): \\ (s,t)\notin E(\cN^1(i))}} \frac {1} {\sigma_{st}} = 2\ {\{\text{number of pairs } (s,t): dist(s,t) = 2\}} .
    $$
    Note, that the number of all shortest paths between any pair of vertices on the distance two 
    $$
    \sumt_{s,t\in V(G):\, dist(s, t) = 2}\sigma_{st} = \sumt_{i\in V(G)}\sumt_{\substack{j\sim i,k\sim i: \\j\not\sim k}} 1=\sumt_{i\in V(G)}d_i(d_i-1)(1-c_i).
    $$  
    This number can be bounded  
    $$
         \sumt_{s,t\in V(G):\, dist(s, t) = 2}\sigma_{st}\leq d_{max}\  \{\text{number of pairs } (s,t): dist(s,t) = 2\}.
    $$
    Therefore, using the relation between average shortest path length and betweenness from~\cite{Lbetw} and the relation from the theorem~\ref{thm:ccws} we obtain
    $$
        L = 1+\frac 1{n(n-1)}\sumt_{i\in V(G)} BC(i)\geq 1+ \frac 1{n(n-1)}\sumt_{i\in V(G)} BC\big(i,\cN^1(i)\big) = 
    $$
    $$
        =\frac {2\,\{\text{number of pairs } (s,t): dist(s,t) = 2\}} {n(n-1)}\geq \frac {2} {n(n-1)d_{max}} \sumt_{s,t\in V(G):\, dist(s, t) = 2}\sigma_{st} = 
    $$
    $$
        = \frac {2} {n(n-1)d_{max}}\sumt_{i\in V(G)}d_i(d_i-1)(1-c_i) = \frac {2} {(n-1)d_{max}} \Big(\E(s)-\E(sc)\Big) = 2\frac {(1-C)\E(s)} {(n-1)d_{max}}.
    $$
\end{proof}

This theorem provides the additional relation to the relation in~\cite{tuzrelL} between the average shortest path length and average clustering coefficient in graphs, but for any simple undirected graph: 

\begin{cor}
     For any simple undirected graph 
    $$
        L\geq 1+ \frac {(1-C_{WS}) \E(s)-\cov(s,c)} {(n-1)d_{max} },
    $$ where $d_{max}$ is the maximum degree.
\end{cor}

%\section{Discussion}

\section{Acknowledgments}
Support from the Basic Research Program of the National Research University Higher School of Economics is gratefully acknowledged.

\end{document}